\documentclass[a4paper,11pt]{article}
\usepackage{pos}
\usepackage{graphics}
\usepackage{wrapfig}
\usepackage{hyperref}

\usepackage{color}
\definecolor{black}{gray}{0}
\definecolor{g40}{gray}{.40}
\definecolor{g90}{gray}{.90}
\definecolor{r1}{rgb}{.75, .10, .10}
\definecolor{b1}{rgb}{.10, .10, .75}
\definecolor{g2}{rgb}{0.0, .60, .20}

\newcommand{\bi}{\begin{itemize}}
\newcommand{\ei}{\end{itemize}}

\title{
The International Lattice Data Grid --- towards FAIR Data
}
\ShortTitle{International Lattice Data Grid}

\author*[a]{Frithjof Karsch}
\author[b]{Hubert Simma}
\author[c]{Tomoteru Yoshie}

\affiliation[a]{Fakult\"at f\"ur Physik, Universit\"at Bielefeld, 33615 Bielefeld, Germany}
\affiliation[b]{Deutsches Elektronen-Synchrotron DESY, Platanenallee 6, 15738 Zeuthen, Germany}
\affiliation[c]{Center for Computational Sciences, University of Tsukuba, Tsukuba 305-8577, Japan}

\emailAdd{karsch@physik.uni-bielefeld.de}
\emailAdd{hubert.simma@desy.de}
\emailAdd{yoshie@ccs.tsukuba.ac.jp}

\abstract{The International Lattice Data Grid (ILDG) is a community-wide
  initiative to realize the sharing of primary data from lattice QCD
  simulations according to the principles of FAIR data. We
  recall the basic concepts of ILDG as a federation of autonomous
  regional grids with common standards for (meta-)data and services,
  and report on current activities, progress, and plans to restore
  and extend the usability of ILDG.
}

\FullConference{
The 39th International Symposium on Lattice Field Theory,\\
8th-13th August, 2022,\\
Rheinische Friedrich-Wilhelms-Universität Bonn, Bonn, Germany
}

\tableofcontents

\begin{document}
\maketitle

\section{Introduction}
20 years ago it had been realized within the lattice QCD community that the 
large data sets needed for their research projects are not only costly 
to generate but would also be highly valuable input for research projects
of other collaborations. The idea of making the gauge field configurations
generated in lattice QCD simulations of various groups available to the wider 
research community has been brought up at the ``$20^{\rm th}$ International Symposium on 
Lattice Field Theory'' (Lattice 2002) in Boston. It has been suggested to establish 
within the lattice community an ''International Lattice Data Grid'' (ILDG), and to
use a common metadata schema for markup together with a metadata catalogue
for searching \cite{Davies:2002mu}.
This idea found much attention and first reports on the organization and implementation of
ILDG have been given at Lattice 2003 and 2004 \cite{Irving:2003uk,Ukawa:2004he,Maynard:2004wg}.
Major achievements in the following years were the definition of a community-wide
agreed metadata schema (QCDml), and the setup of interoperable storage systems and services
\cite{Coddington:2007gz,Yoshie:2008aw,Beckett:2009cb,Maynard:2009szr}.

\begin{figure}[ht]
  \begin{center}
    \includegraphics[width=0.6\linewidth]{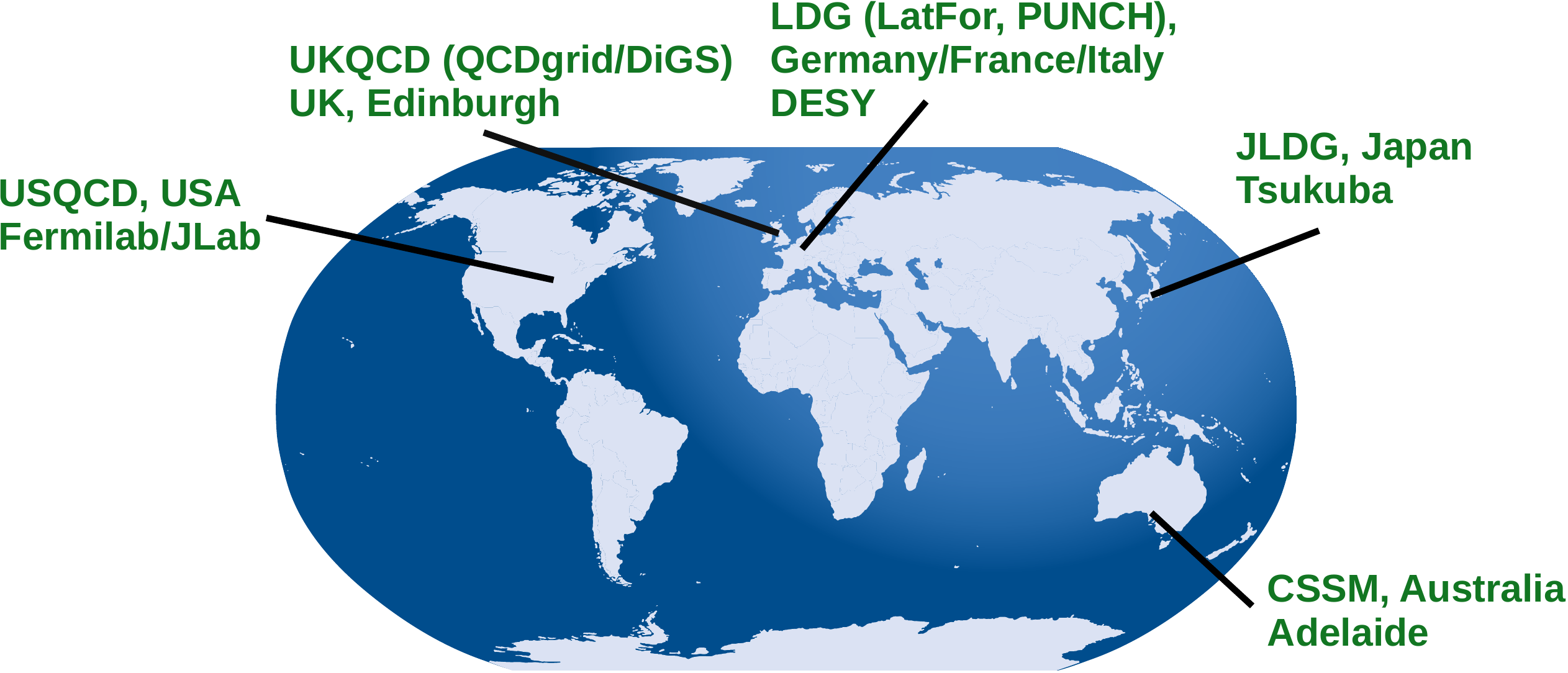}
  \end{center}
\caption{Regional Grids of ILDG}
\label{fig:map}
\end{figure}
 
ILDG started as a ``grid of grids'', i.e. a federation of interoperable
but autonomously operated infrastructures and services. The currently five
``regional grids'' of ILDG are shown in Fig.~\ref{fig:map}. Each of them
is responsible and free to implement and operate the services in its own
(and ILDG-compatible) way. In particular, the regional grids need to acquire
the necessary storage resources for their data.
ILDG only operates a community-wide
user registration and has two working groups \cite{MDWG:ildg,MWWG:ildg} which,
for instance, elaborate the ILDG-wide metadata schemata and the API for the
services of the regional grids. Due to these uniform specifications sharing of
data and searching of metadata is possible in a community-wide and seamless
way, e.g. through simple web interfaces of JLDG or LDG \cite{facetnavi,FacetNaviProc,listing}.
Gauge configurations in the US (with somewhat more specific metadata and access
\cite{USQCD:DMP}) can be found through the DOE data explorer \cite{OSTI}.

The FAIR principles (Findable-Accessible-Interoperable-Reusable) \cite{Wilkinson2016,gofair}
establish essential guiding principles for scientific data management
and are becoming a mandatory requirement by funding organisations.
FAIR principles realized at all levels of the data management, including intermediate and
primary data, is also an important aspect in concepts, like Open Data and Open Science
(see also presentations at this conference \cite{Athenodorou:2022ixd,Bennett:2022klt}).

When bringing to life ILDG, the lattice community already made an important
step towards FAIR data even before these principles were 
explicitly formulated and commonly accepted. During the last years,
however, usability and usage of ILDG has severely
degraded. This has partially been a consequence of the lacking broader uptake
of Grid technologies, which made it increasingly difficult keeping
the infrastructure and the access to it alive. Instead, Cloud
technologies have become the primary choice for realising distributed
data infrastructures due to the success of large commercial public
Cloud providers.

On the other hand, the necessity and interest to share lattice data, at least
(but not only) at the level of our costly raw data, is high. This has become
evident, for instance, in the parallel session on ``Lattice Data'' at this conference
\cite{parallel}. More than ten collaborations indicated interest in making order of
500 ensembles openly accessible through ILDG. These sum up to more than 15 million
configurations and about 5 PB. Sharing such volumes is a challenging task in view
of limited storage resources and person power. 

Therefore, resuming and joining efforts to restore and improve ILDG seems
desirable and timely. Fortunately, ILDG can eventually also leverage on some
recent funding, e.g. within the PUNCH4NFDI project \cite{punch4nfdi}, for
urgently needed software developments to ``go FAIR''. The main objectives for ``ILDG 2.0''
are to improve usability of ILDG and to re-align it with the technological
evolution, e.g. towards Cloud storage.
Ongoing activities include the transition to token-based authentication
(instead of grid certificates), containerization (for easier deployment in
different regional grids) and re-factoring of metadata catalogue and user tools,
as well as revision of the metadata schemata and support for DOI registration
and data publishing. Since some of these topics are not lattice-specific, we
also hope to exploit or create synergies with related research-data
efforts in other fields.

\section{Basic Elements of ILDG}
\subsection{Global Structure and Services}
ILDG is a federation of autonomous ``Regional Grids'' (RG) with a single Virtual
Organization (VO). 
Users can register as members of the VO through a unique registration service
which then provides support for further authentication and access to other
services. The user registration of ILDG is currently still based on authentication
via grid certificates and implemented through a VO Membership Service (VOMS) \cite{VOMS}.

Moreover, ILDG has agreed on community-wide standards and conventions for
(i) a binary storage format of gauge configurations, (ii) a metadata
schema and format used for markup of the data, and (iii) a minimal set of
services (including their API) to be operated by the regional grids.

The VO membership registration service and a web page \cite{ILDG:web}, where the 
specifications (i), (ii), and (iii) can be found, 
are in fact the {\em only} services operated at a global level by ILDG. 

\subsection{Services of the Regional Grids}
The basic services to be provided and operated by each regional grid are
\bi \setlength{\itemsep}{1mm}
\item a Metadata Catalogue (MDC) with an ILDG-compatible application programming
interface (API) to (i) register
  the unique identifiers for the data (i.e. ensembles and configurations)
  provided by the RG, (ii) store the corresponding metadata, and (iii) to
  make it searchable through a powerful query language (Xpath).
\item a File Catalogue (FC) which for each configuration identifier returns
  the list of storage locations (SURLs), where the (possibly replicated)
  configurations can be accessed through standard protocols.
\item storage elements (SE) where the configurations are actually stored.
\item a web page with any further RG-specific information.
\ei
The URLs of these services from each individual regional grid are kept in
a machine actionable file 
on the ILDG web page, e.g. for use by automated client tools or scripts.

The regional grids are responsible for setting up and autonomously
operating these basic services following the ILDG specifications.
In this way, each RG has large freedom in the implementation
choices for services and infrastructure, and vendor locks are avoided.
For instance, the regional grids in Japan (JLDG) \cite{JLDG} and 
Europe (LDG) \cite{LDG} use rather different solutions for data storage:
a global file system (GFARM) in case of JLDG \cite{Amagasa:2015zwb},
and several (distributed) storage elements with SRM interfaces being part
of the Worldwide LHC Computing Grid in case of LDG.

Due to the common API and standard protocols used for the basic services,
every regional grid or even individual persons can develop their own client tools
or additional services -- either according to their own needs and preferences,
or as an offer for community-wide use. Examples include convenient web interfaces
for simple listing \cite{listing} or advanced (meta-)data searches \cite{FacetNaviProc}.

Of course, neither the Virtual Organization ILDG nor the regional grids are usually direct
service providers. Operating the services and, in particular, providing the hardware
for storage elements, relies on the support by the home institutions of the
scientists or on other sources.

\subsection{Organization}
ILDG has two working groups: one for metadata-related topics,
and one for middleware-related topics.
After a phase of inactivity during recent years, the working groups have 
resumed regular online meetings \cite{Meetings} since beginning of 2022 
to discuss status, plans, and ongoing or future activities within the regional
grids and ILDG wide.
Major topics of the Metadata Working Group (MDWG) \cite{MDWG:ildg} currently
include adjustments of the metadata schema \cite{MDWG:doc}
and support for data publishing (see also Sect.~\ref{ss:publishing} below).
The Middleware Working Group (MWWG) \cite{MWWG:ildg}  works on technical aspects
including the development, maintenance, and improvement of services, infrastructure,
and user tools of ILDG and its RGs (see also Sect.~\ref{ss:ltools} below).

The ILDG Board \cite{Board} is composed
of one to two representatives of each RG and the convenors of the working groups.
The board meets several times per year to discuss and decide any major strategic and
organizational matters of ILDG, including the directions for the activities of the
working groups, and the organization of meetings and other outreach activities
targeted to the entire the lattice community (e.g. virtual workshops, or the
the parallel session on ``Lattice Data'' and the ILDG lunch meeting at this conference).

\subsection{Use Cases and Requirements}
Data management and sharing of gauge configurations usually involves two
different user perspectives:``data providers'', who carry out the simulations
and generate the gauge configurations, and  ``data consumers'', who use the
configurations for further processing and analysis.

Moreover, configuration sharing can take place at different levels:
at a collaboration-internal level within one (or, in case of joint
projects, few) collaborations, or at a community-wide level. 
Often, ensembles of gauge configurations which initially are only shared
within a collaboration are later on made publicly available for the
entire lattice community after some embargo time.

The aim of ILDG is to provide a framework that is convenient and beneficial for
all these use cases, not only for data consumers or at a community-wide level.
At a first glance, this is not obvious, because markup of metadata and
packing of configurations according to the schema and format required by ILDG
may seem an extra burden for data providers. However, the information required
for the ILDG metadata is essentially just the kind of metadata which anyhow
is required for any rigorous and high-quality data management and curation,
and which can easily be collected in any well organized data production
workflow. If this information, basically a key-value list, is properly collected
right from the beginning (possibly also in some simple custom format), it is then
trivial to convert it into the ILDG markup. Thus, ILDG can help the data
providers to guide and organize the workflow for data management and sharing
already at the collaboration-internal level in a clear and transparent way, and to
guarantee a high level of data quality (e.g. through the checksum and provenance
information included in the ILDG metadata).

In fact, support for convenient data management and sharing at a collaboration-internal
level has been required and intensively used by several collaborations within JLDG \cite{Amagasa:2015zwb}
or LDG (e.g. ETM or QCDSF) since the beginning of ILDG. Of course, a well-defined and
fine-grained access control is a critical technical requirement for this purpose.
Then, if configuration sharing through ILDG is already put into place at the
collaboration-internal level, making an ensemble eventually community-wide available
simply amounts to toggling a flag in the access control settings at the end of the 
embargo period.

Both at collaboration-internal or community-wide level, a suitable way to
cite ensembles that were used as input data for further analysis is a basic
element of good scientific practice. Moreover, it allows data providers to receive
adequate credits and citations, and to properly acknowledge computing grants. Since
typically many ensembles, e.g. with different lattice spacings and quark masses,
enter into a physics result, citation in form of a lengthy list of ensemble identifiers
of ILDG is often not convenient or practical. Therefore, an additional support or workflow
to publish (sets of) ensembles in a more standard way, including the assignment of
DOIs which then can be conveniently cited, is an important extension of ILDG.

\section{ILDG goes FAIR}
Keeping in mind that ILDG started more than a decade before the FAIR principles
were actually formulated in \cite{Wilkinson2016}, it is instructive to ask how
well the concepts and implementation of ILDG already are compliant with these
FAIR principles and where adjustments and extensions are necessary for ILDG 2.0.
It is also important to recall that \cite{Wilkinson2016} describes
guiding principles for scientific data management, not an implementation.

In the following we only consider the ten main criteria of \cite{Wilkinson2016}
which are shown in the text boxes below. A more in-depth analysis, e.g.
according to the FAIR Data Maturity Model \cite{FDMM}, would be beyond the
scope of this contribution.

\subsection{Findable}
\begin{center}
\fcolorbox{r1}{g90}{\parbox{0.85\linewidth}{
    \begin{itemize} \setlength{\itemsep}{0mm}
    \item[F1] globally unique and persistent ID assigned to (meta-)data
    \item[F2] data described with rich metadata 
    \item[F3] metadata includes data ID of the data
    \item[F4] (meta-)data is registered or indexed in a searchable resource
    \end{itemize}
}}
\end{center}
In ILDG, an identifier of the form
\begin{equation}
    {\tt lfn:}//\langle rg\rangle/\langle collab\rangle/\langle proj\rangle/\ldots
    \label{e:lfn}
\end{equation}
or
\begin{equation}
  {\tt mc:}//\langle rg\rangle/\langle collab\rangle/\langle proj\rangle/\ldots
  \label{e:mc}
\end{equation}
is assigned to each gauge configuration and to each ensemble, respectively.
Including the names of the regional grid ($rg$), collaboration ($collab$), and
project ($proj$), together with a suitable convention at the level of each
project, guarantees global uniqueness (F1). Persistence of the identifiers
is realized in practice for at least the lifetime of the (meta-)data, which
in case of the earliest ensembles uploaded in LDG is already 16 years.

Gauge configurations and ensembles are described in ILDG by metadata
with a rich, hierarchical, and flexible structure (F2) which is concisely
defined in terms of two corresponding metadata schemata.
As required by (F3), the metadata of ensembles
and configurations 
also include the unique identifiers (\ref{e:lfn}) and (\ref{e:mc}). The
corresponding metadata elements are called {\tt dataLFN} (data logical
file name) and {\tt markovChainURI}, respectively.

Since the data files of the configurations are large, they are usually stored
in a distributed manner on different systems. In contrast, the metadata is
much smaller and kept separately from the data in order to allow efficient
searching. The metadata in ILDG is registered and stored in the central
Metadata Catalogue (MDC) of each regional grid. The MDC supports complex
queries on the content of the metadata and returns the list of matching
identifiers (\ref{e:lfn}) or (\ref{e:mc}). Thus, the MDC of each regional
grid is an essential building block of ILDG to satisfy the requirement (F4).

\subsection{Accessible}
\begin{center}
  \fcolorbox{r1}{g90}{\parbox{0.85\linewidth}{
    \begin{itemize} \setlength{\itemsep}{0mm}
      \item[A1~~] (meta-)data retrievable by ID using standardized protocols 
      \item[A1.1] protocol is open, free, and universally implementable 
      \item[A1.2] protocol allows authentication/authorization procedure where necessary 
      \item[A2~~] metadata accessible even when data is no longer available
    \end{itemize}
}}
\end{center}
The Metadata Catalogue in ILDG is a web service (not just a web page) and can
be accessed via the standard HTTP(S) protocol and a well-defined API contract.
As required by (F4) and (A1), the MDC supports searching through
the configuration and ensemble metadata (by Xpath queries), as well as retrieving the {\em metadata}
itself by their {\tt dataLFN} or {\tt markovChainURI}, respectively. Since all metadata in
ILDG are public, authentication or authorization is only used for write access (A1.2).

The actual {\em data}, i.e. the gauge configurations, are retrievable by their {\tt dataLFN}
(as required by A1) with the help of the File Catalogue (FC), which logically is
a separate service in ILDG, but can be also be considered as part of the MDC: first,
a query to the FC with a the desired {\tt dataLFN} returns the storage URL (SURL)
of the configuration (or a list of SURLs if data is stored redundantly). The SURL
includes not only the storage location (address and path), but also a standard
protocol (e.g. GSIFTP, HTTP, or SRM) to access the corresponding storage system (A1.1).
The data can then be downloaded by standard clients, like {\tt globus-url-copy} or
the {\tt gfal2} library.
By providing suitable client tools (e.g. scripts), users can conveniently carry
out all steps 
through a single command. 

The FAIR requirement (A2) is remarkable because it suggests that metadata is to be
considered at least as important as the data itself. In ILDG this can and needs to
be taken into account, for instance, by frequent (daily) automatic backups of the
MDC data.

The API of the MDC and FC service for ILDG are defined by the Middleware Working Group.
Using the same API in each regional grid ensures compliance with (A1.1) and
interoperability\footnote{
Interoperability is the ability of {\em data or tools} from non-cooperating 
resources to integrate or to work together with minimal effort \cite{Wilkinson2016}.
See next sub-section for interoperability at the (meta-)data level.}
of their services. 
While the current API is based on the
Simple Object Access Protocol (SOAP), the transition to a Representational State
Transfer (REST) style is in preparation.

In practice, because of IT security concerns, e.g. in data centers, most of
the storage systems used by our community also require authentication (A1.2)
--- even for read access (at least for large amounts of data and fast protocols).
Therefore, a global user registration in a VO is essential for providing
the authentication and authorization services necessary for seamless and
community-wide configuration sharing.

\begin{figure}[ht]
  \begin{center}
    \includegraphics[width=0.7\linewidth]{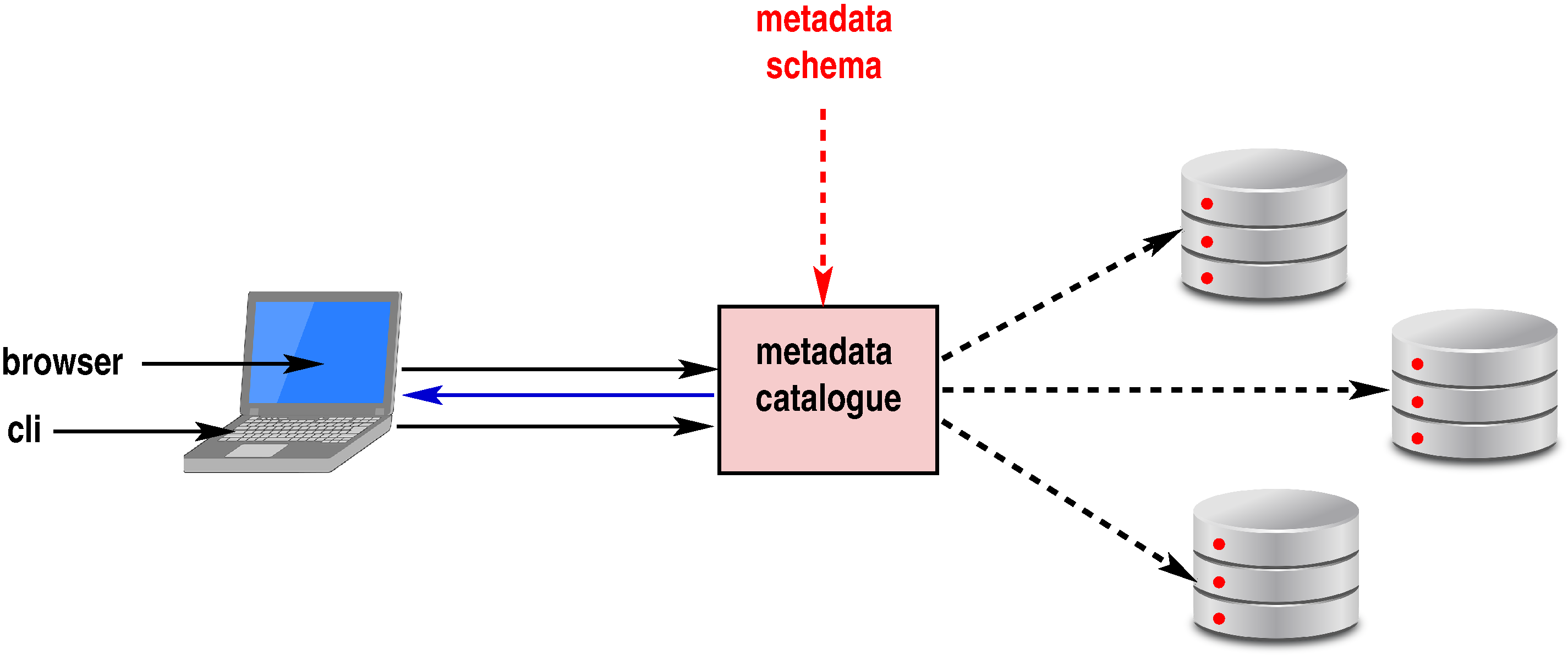}
  \end{center}
\caption{Distributed storage in a regional grid with a central Metadata Catalogue}
\label{fig:mdc-schema}
\end{figure}

The simplified structure of a regional grid in ILDG with findable and accessible
data is illustrated in Fig.~\ref{fig:mdc-schema} (without explicitly showing the
FC queries and authorization flow). From an abstract point of view, ILDG just
implements a database system on a distributed infrastructure by using technologies
and concepts of the World Wide Web. The basic entities are configurations and
ensembles labeled by unique identifiers. The metadata describing e.g. the physical or
administrative properties of the configurations and ensembles --- or in case
of configurations also the actual data itself --- are logically just attributes
associated with these identifiers.

\subsection{Interoperable}
\begin{center}
  \fcolorbox{r1}{g90}{\parbox{0.85\linewidth}{
    \begin{itemize} \setlength{\itemsep}{0mm}
    \item[I1] (meta-)data use a formal, accessible, shared, and broadly applicable 
      language 
    \item[I2] (meta-)data use vocabularies that follow FAIR principles
    \item[I3] (meta-)data include qualified references to other (meta-)data
    \end{itemize}
}}
\end{center}

Data and metadata in ILDG satisfy  (I1)--(I3) by consistently
using well-defined storage and markup formats which are an integral part
of the ILDG specifications.

Metadata is stored in XML format and must validate against the XSD definition
of the corresponding schema. In addition to the XSD specification, further
documentation of all elements of the schema can be found on the web page of
the ILDG Metadata Working Group \cite{MDWG:doc}.

Gauge configurations are stored in binary records with specified order
and format (big-endian, 32- or 64-bit IEEE floating-point). Moreover,
ILDG specifies a very simple packaging of the data record together with
additional metadata (lattice geometry and {\tt dataLFN}, plus optional
user-defined metadata) into a single file.
  
\subsection{Reusable}
\begin{center}
  \fcolorbox{r1}{g90}{\parbox{0.85\linewidth}{
    \begin{itemize} \setlength{\itemsep}{0mm}
    \item[R1~~] (meta-)data are richly described with plurality of accurate and relevant\\ attributes
    \item[R1.1] (meta-)data are released with clear and accessible data usage license
    \item[R1.2] (meta-)data are associated with detailed provenance
    \item[R1.3] (meta-)data meet domain-relevant community standards
    \end{itemize}
}}
\end{center}

At the highest level, the two metadata schemata of ILDG are grouped into sub-trees,
like {\tt management}, {\tt physics}, {\tt algorithm}, {\tt implementation}, or
{\tt markovStep}. These are further structured to specify, e.g. lattice geometry,
type and parameters of the action and the simulation algorithm, simulation code
and hardware, further provenance information (R1.2), checksum, and plaquette value,
etc.,

The two metadata schemata of ILDG seem to meet very well the requirement (R1), in
particular (R1.2) and (R1.3). However, a clear data usage license, as required by
(R1.1), is missing and must become mandatory in the future, also in order to fully
support a data publishing process.

Reusability is also related to the important concepts of reproducibility
in Open Science (see presentation by
E.~Bennett at this conference \cite{Bennett:2022klt}).

\section{Towards ILDG 2.0}
Main objective of the ongoing reactivation and restart of ILDG, which we coin
``ILDG 2.0'', is to set up a fully operational, maintainable, FAIR-compliant
and user-friendly framework -- for data providers as well as data consumers.
Improvements needed for this purpose include
\bi
\item development and maintenance of user tools (e.g. CLI clients), together with
  up-to-date user documentation
\item support for data publishing and DOI registration, in order to make the data citable
\item revision of the metadata schemata and data formats (e.g. HDF5)
\item support for finding the data through convenient web interfaces or standard search engines
\item technology updates, including a VO-wide token-based authentication and authorization
\ei

\subsection{User Tools}
\label{ss:ltools}
Due to the common API and standard protocols used for the basic services,
regional grids or even individual persons are able to develop client tools
or additional services. However, coordinating and sharing such efforts in
a more coherent way is needed to improve the usability and quality of ILDG
at a community-wide level.  

For the use cases discussed in Sect. 2.4, only a few high-level operations are
needed and can be performed by command-line tools, like\footnote{The command names
  used here are fictitious, but inspired by the {\tt ltools} commands that were
  provided and used by LDG in the past.
}
\bi
\item {\tt lfind} to query the MDC of a regional grid (or of all of them)
  for ensembles or configurations which have specific properties. Since the MDC
  supports powerful Xpath queries, one can search for any combination of properties 
  contained in the metadata. For instance, one might want to search for ensembles
  which have a certain action, number of flavors, lattice size, value of physics
  or algorithm parameters, etc. In the same way, one might search e.g. for all
  configurations that were generated by a specific collaboration or code, on a
  specific machine or in some period of time.

\item {\tt lget} with a list of identifiers ({\tt dataLFN} or {\tt markovChainURI})
  as argument to either download the corresponding metadata and/or (depending on
  command-line options) the configuration data itself. In the later case, {\tt lget}
  should implicitly perform look-up in the FC, authentication (if needed), and
  verification of the checksum.

\item {\tt linit} is a command needed by data providers or project managers to
  register a new ensemble ID with its metadata, and to set additional attributes,
  like collaboration and project name, or access rights.

\item {\tt lput} can then be used to upload configuration metadata and data.
  In the later case, the storage element might be selected by default or explicitly
  specified,  and {\tt lput} implicitly carries out registration of the storage URL in the FC,
  as well as the required authentication and authorization flow

\ei

Also the effort to prepare metadata and data for uploading, can be strongly
reduced by clearly separating the task into generic and project-specific steps.
Project- or collaboration-specific steps usually need to be set up only once.
This setup phase amounts to preparing the following prerequisites:
a workflow to generate gauge configuration data in ILDG format (directly
or by a conversion tool if the simulation code uses a custom format);
and creation of a template for the metadata. Such a template can contain
all the information, which typically is invariant for an entire set of ensembles
(like the description of the action or the algorithm), together with place holders
for a limited set of variable values, which can usually be determined in a simple
and automatic way, e.g. by trivial scripts or commands to extract them from log
files, etc.

Then, the repeatedly required steps for actually uploading,
namely packing the configurations and generating the metadata,
can be handled automatically by two simple and generic
tools. A common and shared implementation of these two
additional tools can thus greatly simplify the upload in ILDG.
  
Many such convenient tools have been created, either within regional grids or by
individual users, in the early days of ILDG. However, due the lack of maintenance,
documentation, and porting to newer software or operating system versions, most of
them have become unavailable. This can be cured by a careful and well documented
re-design of these user tools. They should have minimal dependencies on any specialized
software (e.g. by just using scripts, and standard commands, like {\tt curl} and
{\tt globus-url-copy}), and/or be packaged into containers (like singularity/apptainer)
for convenient deployment on different user platforms (including systems where users
do not have root privileges).

\subsection{Search Interfaces}
A simple command-line tool for searching (like {\tt lfind}) needs as argument
a single string which represents the search criteria as an Xpath query. However,
properly formulating the query strings for complex searches is not completely
straightforward and requires some knowledge of Xpath syntax and of the metadata
schema. In addition --- and as an alternative --- to command-line tools it would,
therefore, be important to also provide convenient (web) interfaces, where queries
can be formulated at a somewhat higher level, e.g. through a user-friendly
selection menu or form similar to INSPIRE-HEP \cite{inspirehep}.

Since the services (like MDC and FC) of all regional grids have the same API,
any implementation of such web interfaces can easily be used at an ILDG-wide
level. Extensions of existing attempts, like \cite{FacetNaviProc,facetnavi} or the
simple listings \cite{listing}, can provide a natural and promising starting point.

\subsection{Revision of Metadata Schemata and Data Format}
\label{ss:md}
Creating a community-wide agreed metadata schemata at the beginning of ILDG was
a challenging task for Metadata Working Group and represents an important
achievement of ILDG. The two schemata have now been in use for more than a
decade with only minimal changes. Clearly further developments and
adjustments are necessary for ILDG 2.0. 

For instance, some elements of the markup schema, in particular the {\tt action}
part in the ensemble metadata, seem to be rather complicated and rigid.
This shall be fixed in future revisions by allowing e.g. references to
external literature or documents, as well as optional glossaries or annotations.

Some metadata elements also need to be adjusted to enable
or simplify markup and uploading of data from simulations with new features.
Such minor adjustments of the schemata will most efficiently be handled
in a use-case driven way: several collaborations, which actively
contribute to the reactivation efforts of ILDG, are preparing major
uploads. Thus, difficulties or concrete solution proposals, which might
arise in this phase, can be directly taken up and finalized by the working
groups.

On the other hand, major extensions of the metadata schema, or an additional schema, might
become necessary in order to fully support a proper data publishing process. 

For the configuration data, also an alternative packaging of the records
according to the widely used HDF5 format and with multiple configurations
per file is being discussed as a desirable extension.

\subsection{Support for Data Publishing}
\label{ss:publishing}
Possibilities to register persistent identifiers, like DOI, have been
discussed in ILDG since a long time. Finding an ILDG-wide solution
or at least best practice for this purpose is clearly among the
important objectives for ILDG 2.0. Being able to register DOIs for
ensembles or sets of ensembles would make citation of these data in
publications much more practical and attractive. In turn, this would
help to motivate and reward data providers who share and upload
their configurations.

A closer look at the publishing process, as it is well established
for literature, reveals that an analogous process is also desirable
and necessary for data \cite{roadmap}. It is not sufficient to just make the data
publicly availability on some computer in the internet and to register
a DOI. For instance, specific information, i.e. additional metadata,
needs to be provided, and a landing page must be generated and hosted.
This is not only required for the DOI registration itself, but also
for making the data findable from non lattice-specific search engines.
Open issues are also the persistence requirements of data and metadata,
or explicit support for metadata harvesting, e.g. through an OAI-PMH
interface and/or arrangements with portals, like INSPIRE-HEP \cite{inspirehep}. 

Interesting solutions for DOI registration and the entire publishing
process have already been explored by JLDG and USQCD, see \cite{JLDG:doi}
and \cite{OSTI}. Also services, like Zenodo
\cite{Zenodo}, seem an attractive option. The Metadata Working Group
is now further investigating these approaches in order to find out
whether an ILDG-wide solution, or at least an ILDG-wide support and
best practice for (possibly different) solutions within the regional
grids, can be realized. For instance, already a community-wide schema
for collecting the metadata, which is required for DOI registration
and for the generation of meaningful landing pages for lattice data, would
be an important step forward.

\subsection{Technology Upgrades}
Several services and building blocks of ILDG have been in use for more
than a decade and a transition or upgrade to up-to-date technology is
overdue.

A key element is the ILDG-wide user management and authentication,
which is currently based on X.509 certificates (in parts with VOMS 
extensions). Switching to a token-based solution (based on OIDC or 
SAML) is necessary to ensure alignment with technologies commonly 
used in the Cloud. This will also free the users from the sometimes 
difficult procedure to obtain a grid certificate.

Among the various alternatives that have been investigated by the
Middleware Working Group, the INDIGO IAM \cite{INDIGO} appears as
the most suitable and promising solution. The deployment of a dedicated
instance of this service at CNAF in Italy will be further explored
and might become a central building block for ILDG 2.0.

Further essential elements of ILDG are the Metadata and File Catalogue
of each regional grid. The MDC implementation, which has been in continuous
operation in LDG since 2008, has recently been re-built 
and containerized. Thus, it can now be easily deployed also in other
regional grids and save redundant development efforts. Integration of
the FC functionality into the MDC is currently in progress and can
remove the need for a separate service. This implementation of the MDC also
provides an additional attribute service for fine-grained access control,
which is essential for collaboration-internal configuration sharing
during initial embargo periods.

Since token-based authentication and authorization, as well as a REST API
for MDC and FC are expected to become an ILDG 2.0 standard, a complete
re-factoring of the MDC and FC is planned for 2023.

\section{Outlook}
Taking up and evolving the ILDG initiative is a natural, efficient, and timely
step to face the data management challenge in Lattice QCD in a way which
is fully compliant with FAIR principles and modern data repository standards.
The basic concepts of ILDG as a federated system of storage infrastructures and
data services are sill and more than ever valid.
While some of the components of ILDG and its regional grids are
still running after more than a decade of stable operations, e.g.
storage elements and the metadata repositories of JLDG and LDG,
modernizations and technological updates are necessary and timely. For
instance, the transition to token-based authentication and authorization
mechanisms is perfectly in line with the ongoing developments of large
HEP experiments towards Cloud technologies. Some of the necessary developments
in ILDG also need dedicated software expertise, which eventually has become
available only thanks to recent funding, e.g. in the context of PUNCH4NFDI.

Two crucial elements for the reactivation and modernization of ILDG
are the services for the VO-wide user registration and authentication,
and the Metadata and File Catalogues. With the completed rebuild and
containerization of the MDC and FC of LDG, these can also
be deployed in other regional grids. Thus, we expect to achieve fully
operational and ILDG-compliant services, at least for JLDG and LDG
and possibly other regional grids within the next months. Larger
uploads e.g. of configurations from ETMC, HotQCD, and CLS, are then
planned, and will serve to improve and test the setup.

The volume of the configurations to be eventually uploaded is of the
order of several PB. Since neither ILDG nor the regional grids have
usually own storage resources, additional storage space may become
a concern. It should therefore be emphasized that lattice collaborations
also need to apply for additional storage resources at their home
institutions or through other sources. Joining efforts in the framework
ILDG is likely to be an advantage for this purpose.

The further evolution of ILDG as outlined in this contribution, in
particular, also development and maintenance of user tools or updated
documentation, relies to a large extent on contributions from the
user community. Therefore, joint efforts and active participation,
e.g. in the working groups is essential and highly welcome.

\section{Acknowledgments}
ILDG is an effort and achievement of the worldwide lattice community, and certainly
not properly represented by the authors of this proceedings. Many thanks are
due to the valuable contributions and help from numerous persons and institutions.
We explicitly like to acknowledge (and excuse for any unintentional omissions):

\bi
\vspace*{-1mm}
\setlength{\itemsep}{0mm}
\item all organizers of this conference and C. DeTar for chairing the ILDG lunch meeting
\item members of the ILDG board:\\
  B.~Blossier,
  R.~Edwards,
  W.~Kamleh,
  Y.~Kuramashi,
  D.~Leinweber,
  A.~Portelli,
  F.~Di Renzo,
  J.~Simone
\item members of and contributors to the Metadata and Middleware Working Groups:\\
  T.~Amagasa,
  G.~Andronico,
  C.~DeTar,
  J.~Hetrick,
  B.~Joo,
  O.~Kaczmarek,
  G.~Koutsou,
  H.~Matsufuru,
  C.~McNeile,
  Y.~Nakamura,
  M.~Di Pierro,
  D.~Pleiter,
  M.~Sato,
  C.~Urbach,
  O.~Witzel,
  J.~Zanotti
\item essential contributions in the framework of PUNCH4NFDI by\\
  B.~Bheemalingappa Sagar,
  D.~Clarke,
  N.~Meyer,
  J.~Simeth,
  and C.~Schmidt-Sonntag 
\item C.~Allton, A.~Athenodoru, E.~Bennett, C.~K\"orber, and many others for helpful discussions and comments
\item lattice participants of PUNCH4NFDI:
  G.~Bali,
  S.~Collins,
  S.~Krieg,
  T.~Wettig,
  C.~Urbach
\item O.~Freyermuth, C.~Wissing, and many other members of PUNCH4NFDI for valuable discussions and comments
\item support and advice by the IT groups at Bielefeld, DESY, J\"ulich, and Regensburg, in particular,
  P.~Bolle, T.~Beermann, P.~Fuhrmann, A.~Gellrich, A.~Haupt, H.~Hess, M.~Klappenbach, K.~Leffhalm, C.~Manzano, 
  S.~Solbrig, C.~Voss, G.~Waschk, T.~Wetzel
\item the INDIGO IAM team, in particular, F.~Agostini and F.~Giacomini
\item services and support for ILDG by CNAF, CSSM, DESY, EGI, EPCC, FZJ, I3N2P, JaLC, NERSC, OSTI, and Tsukuba University
\ei
This work was in part supported by DFG fund "NFDI 39/1" for the PUNCH4NFDI consortium. 
FK was also supported by the DFG Collaborative Research Centre 315477589-TRR 211, ”Strong interaction matter under extreme conditions”. 
HS acknowledges funding from the European Union’s Horizon 2020 research and
innovation programme under the Marie Skłodowska-Curie grant agreement No. 813942 (ITN EuroPLEx).
TY would like to thank the CCS, University of Tsukuba for continuous financial support for ILDG activities.

\bibliographystyle{JHEP}
\bibliography{ildg.bib}

\end{document}